\documentclass[a4paper,11pt]{article}
\pdfoutput=1 

\usepackage{jheppub} 

\usepackage{physics}

\usepackage[T1]{fontenc} 

\title{\boldmath Giant Gravitons Intersecting at Angles from Integrable Spin Chains}

\author[a]{Adolfo Holguin}

\affiliation[a]{University of California Santa Barbara, Department of Physics}

\emailAdd{adolfoholguin@physics.ucsb.edu}

\abstract{We study integrable non-diagonal open boundary conditions for spin chains arising from holographic gauge theories. Their dual description is in terms of open strings stretching between giant gravitons intersecting at arbitrary angles inside spheres or projective spaces. By studying properties of their ground and low-lying states we reproduce the expected spectrum near the intersections. Unlike the case of intersecting D-branes in flat space, intersecting giant gravitons related by $SU(N)$ rotations do not always preserve an additional supersymmetry. We quantify this lack of enhancement by whether or not there exists a simple ferromagnetic ground state for the open spin chain. }

\begin{document} 
\maketitle
\flushbottom

\section{Introduction}
D-branes are an important class of non-perturbative BPS states that preserve one-half of the supersymmetries of superstring theories \cite{Polchinski:1995mt}. For D-branes intersecting in flat space, the spectrum of open strings stretching between them enjoys an enhanced supersymmetry and the spectrum of worldsheet momenta encodes the angle at which the branes intersect \cite{Berkooz:1996km}. The spectrum of open strings between D-branes in curved spaces is an active area of study \cite{Ooguri:1996ck,Blumenhagen:2005mu}. The AdS/CFT correspondence provides an avenue for studying two important problems: gauge theory at strong coupling and quantum gravity in anti-de Sitter space \cite{Maldacena:1997re}.  In the past two decades, integrability has proven to be a powerful tool for studying the spectrum of  the $SU(N)$ $\mathcal{N}=4$ super Yang-Mills theory at large $N$ (or equivalently free type IIB superstrings on $AdS_5\times S^5$), due to the fact that the mixing problem for single trace operators can be recast as a spectral problem for an integrable Hamiltonian acting on a special type of spin chain \cite{Beisert:2010jr}. These operators correspond to perturbative excitations around the background and they have dimensions of order one. Operators with dimensions of order $N$ and larger describe non-perturbative objects of the theory, namely the giant gravitons of  McGreevy, Susskind, and Toumbas \cite{McGreevy:2000cw}. These giant gravitons are D-branes wrapping compact cycles of the spacetime which carry an extremal amount of charge. In their spin chain description, these operators serve as boundary conditions for the spin chain and remarkably, the giant gravitons with maximal charge $J=N$ provide a class of integrable boundary conditions \cite{Berenstein:2005vf}.

Since the giant gravitons are localized inside the bulk $AdS_5\times S^5$ geometry, the open strings stretching between them are able to probe the bulk geometry on which said strings move \cite{Berenstein:2003ah}. For configurations of non-intersecting giant gravitons, the energy of the open strings between them encodes the geometric distance between the branes. Similarly, the spectrum of open strings between intersecting giant gravitons is sensitive to the angle at which the D-branes intersect inside the geometry \cite{Berkooz:1996km}. 

In this paper we study simple configurations of interesecting maximal giant gravitons in $AdS_5 \times S^5$ or $AdS_4\times \mathbb{CP}^3$ at generic angles. In section 2 we review the spin chain description of ``long'' operators in $\mathcal{N}=4$ super Yang-Mills and the $\mathcal{N}=6$ ABJM theory at weak coupling. In section 3 we describe non-diagonal boundary conditions for some simple sectors of the spin chain and give their geometric interpretations. In section 4 we compute various properties of the ground states and one-defect states of the resulting spin chains. For the $\mathcal{N}=4$ case we are able to find explicit forms for the ground state and one-magnon eigenstates, but for the ABJM theory we resort to a coherent state variational ansatz.  In the $AdS_5 \times S^5$ case, we illustrate how boundary conditions which support a ferromagnetic ground state agree qualitatively with the same analysis in flat space, with suitable modifications coming from the curved background. We argue that configurations like these generically preserve one half of the supersymmetries near the intersection. For  $AdS_4\times \mathbb{CP}^3$, we find that the generic intersection does not yield a ferromagnetic groundstate, and argue that this describes a massive quarter BPS state of the \textit{centrally extended} supersymmetry algebra, whose central charge encodes the angle of the intersection. 

\section{Integrable Spin Chains for Holography and Strings}

\subsection{The $\mathcal{N}=4$ SYM Spin Chain}

It is well known now that certain problems in gauge theory can be recast as spectral problems for a Hamiltonian acting on a spin chain. The canonical example is that of the dilatation operator acting on single-trace operators of the $\mathcal{N}=4$ superconformal Yang-Mills theory in the planar limit. For operators made out of $L$ scalar fields, the one-loop dilatation operator is given by
\begin{equation}
    D_1= \frac{\lambda}{8\pi^2} \sum_{i=0}^{L}\left(1- P_{i,i+1}+\frac{1}{2} K_{i,i+1} \right),
\end{equation}
where $P_{i,j}$ and $K_{i,j}$ are the permutation and trace operators acting on a tensor product of fundamental representations of $SO(6)$ \cite{Minahan:2002ve}. Each site of the spin chain is described by a linear combination of letters inside the trace in the operator. In general, the letters can be chosen to be any of the fields of the $\mathcal{N}=4$ super Yang-Mills theory:
\begin{equation}
    \texttt{letters}= \{ D_\mu, \Psi_I, X, Y, Z, \dots\}.
\end{equation}
In the $SO(6)$ sector, we only use the scalar fields $X,Y,Z$ and their complex conjugates.  These can be assembled into words by matrix multiplication, e.g.
\begin{equation}
     \texttt{words}= \{XZZ,  XZZY, XXXYZX,\dots \},
\end{equation}
For more general sectors, these words can include fermions and covariant derivatives. These words then need to be contracted appropriately to create gauge invariant operators. One simple way of doing this is to take the trace of a word of length $L$. This can be thought of as a spin chain of size $L$ with periodic boundary conditions.
Roughly speaking, the permutation operator appears due to Wick contractions of the operator with the $F$-term commutator interactions $|[X,Y]|^2$, while the trace comes from contractions with the $D$-terms $|[X,\Bar{X}]+[Y,\Bar{Y}]+[Z,\Bar{Z}]|^2$. The supersymmetry of the theory can be used to fix the coefficients in front of each operator. With this particular choice of coefficients, the dilatation operator is mapped to an integrable Hamiltonian; it is expected that this continues to hold to all orders in perturbation theory in the planar limit. On the other side of the duality, these spin chains describe the spectrum of a free closed string of the type IIB theory on $AdS_5 \times S^5$. The spectrum of solitons of the classical string matches remarkably well with the defects of the spin chain, with both sides of the duality having a non-relativistic dispersion relation \cite{Beisert:2005tm}:
\begin{equation}\label{dispersion}
    \Delta-J= \sqrt{Q^2+ \frac{\lambda}{\pi^2} \sin^2 \frac{p}{2}}.
\end{equation}
Here $J$ and $Q$ are certain angular momentum quantum numbers of the string and $\lambda$ is the 't Hooft coupling. The momentum $p$ acts as a central charge of the theory; its vanishing on physical states describes the usual level matching conditions that physical closed string states must satisfy. This central charge also encodes information about the geometric position of the string in the bulk spacetime. Since the closed string states are neutral under this charge, their position in the spacetime is not a well-defined gauge invariant observable.

String states that do carry this central charge must be coupled to appropriate sources of this charge in order to preserve gauge invariance: these are open strings ending on D-branes.  Understanding the spectrum of excitations of D-branes in curved spacetimes remains a key problem in understanding non-perturbative aspects of string theories. A particularly simple class of D-branes in the $AdS_5 \times S^5$ geometry are the so-called giant gravitons, which are D3-branes wrapping three-sphere cycles in either $AdS_5$ or $S^5$. The giant gravitons that stretch in the five-sphere directions are described in the gauge theory by (sub)determinant operators \cite{Balasubramanian:2001nh,Corley:2001zk}:
\begin{equation}
\begin{aligned}
          \mathcal{O}_l&= \text{det}_l X, \qquad 1\leq l \leq N,\\
     \text{det}_l  X  &= \frac{1}{l!}\epsilon_{i_1 \dots i_l a_{l+1}\dots a_N} \epsilon^{j_1 \dots j_l a_{l+1}\dots a_N} X_{j_1}^{i_1}\dots X_{j_l}^{i_l}.
\end{aligned}
\end{equation}
These serve as a boundaries to which words in the fields of $\mathcal{N}=4$ SYM can attach. When $l=N$, the associated boundary condition is integrable in the planar limit and can be described by introducing an auxiliary background magnetic field at the boundaries \cite{Berenstein:2005vf}. For a more general discussion of integrability and open boundaries in the context of $AdS/CFT$ see \cite{Zoubos:2010kh}. In the language of the $SO(6)$ spin chain, this involves the introduction of boundary projection operators:
\begin{equation}
\begin{aligned}
        H_{\text{boundary}}&= \lambda \left(Q^{X}_0(1- P_{0,1}) Q^{X}_0+Q^{X}_{L+1}(1- P_{L,L+1}) Q^{X}_{L+1}\right), \\
         Q^{X}&= 1-\ket{X}\bra{X}.
\end{aligned}
\end{equation}
This interaction effectively makes states with $\ket{X}$ in the first and last sites of the spin chain non-dynamical, while allowing the boundary sites to be polarized in the remaining directions. Operators attaching to subdeterminants with $l<N$ are described by more complicated \textit{dynamical} spin chains whose number of sites is not well-defined \cite{Beisert:2003ys,Berenstein:2005fa}. Since these boundary conditions do not conserve the number of defects in the chain, they are not expected to be solvable by a Bethe ansatz.

\subsection{ABJM Spin Chain}
Another example of an integrable spin chain arising from a gauge theory is the $OSp(2,2|6)$ ABJM spin chain. Unlike the $\mathcal{N}=4$ SYM spin chain, the ABJM spin chain involves next-to-nearest neighbor interactions. When restricted to the $SU(4)$ sector, the Hamiltonian at two loops takes the form \cite{Minahan:2008hf}
\begin{equation}
    H_{SU(4)}= \frac{\lambda^2}{8 \pi^2} \sum_{l=1}^{2L}\left(2- 2P_{l, l+2}+ P_{l,l+2}K_{l, l+1} + K_{l,l+1} P_{l,l+2}\right).
\end{equation}
Due to the nature of the bifundamental gauge group, the spin chain must be an alternating chain with fundamental and anti-fundamental representations sitting at odd and even sites, respectively. The corresponding operators are strings of letters alternating between $Y^A=(A_1,A_2, B_1^\dagger, B_2^\dagger )$ and its complex conjugate $Y_A^\dagger$. The defects of this spin chain are also expected to follow the non-relativistic dispersion relation
\begin{equation}
    \Delta-J = \sqrt{Q^2+ 4 h^2(\lambda) \sin^2 \frac{p}{2}},
\end{equation}
where again $J$ and $Q$ are angular momentum quantum numbers associated to the string. Another important difference that arises from the alternating nature of the spin chain is that the allowed boundary conditions are richer than those of the $\mathcal{N}=4$ spin chain. The operators corresponding to giant gravitons in this case are dibaryon determinants:
\begin{equation}
   \mathcal{O}= \det(Y_I Y^\dagger_J).
\end{equation}
Since these determinants factorize, they describe a non-trivial cycle corresponding to a pair of $\mathbb{CP}^2$'s intersecting at a $\mathbb{CP}^1$ inside $\mathbb{CP}^3$. Because of this, the strings can start and end at different parts of this cycle. If one concentrates on a holomorphic $SU(2)\times SU(2)$ sector ($K_{i,i+1}=0$), the spin chain factorizes into a pair of non-interacting $XXX$ Heisenberg chains. The boundary interactions associated with the operator $\det(A_1 B_1)$ are given by~\cite{Chen:2018sbp}
\begin{equation}
\begin{aligned}
     H&= \frac{\lambda^2}{8 \pi^2}\sum_{l=2}^{2L-3}\left(1- P_{l, l+2}\right)\\
     &+\frac{\lambda^2}{8 \pi^2}Q^{A_1}_1\left(1- P_{1,3}\right)Q^{A_1}_1Q^{B_1}_{2L}\\
     &\frac{\lambda^2}{8 \pi^2} Q^{B_1}_{2L}\left(1- P_{2L-2,2L}\right)Q^{A_1}_1Q^{B_1}_{2L}.
\end{aligned}
\end{equation}

\section{Non-diagonal boundaries: Giant gravitons intersecting  at angles}

\subsection{$\mathcal{N}=4$ SYM}

In general we are not restricted to have a word ending on a single determinant, i.e. an open string starting and ending on the same D-brane, or on a separate determinant made out of the same letters. One can for instance consider two different determinant operators made out of a $X$ and $Y$, two of the three complex scalar fields of the theory:
\begin{equation}
\begin{aligned}
\mathcal{O}_0 &= \det \left(\alpha X +\beta Y \right),\\
\mathcal{O}_{L+1} &= \det \mathcal{R}_\theta\left(\alpha X +\beta Y \right).\\
\end{aligned}
\end{equation}
Here $\mathcal{R}_\theta$ is an $SU(2)$ rotation matrix whose eigenvalues are $e^{\pm i \theta}$. Without loss of generality, we can choose our basis of fields $X,Y$ to correspond to the eigenvectors of the matrix $\mathcal{R}_\theta$. Geometrically, this defines a set of two complex lines sitting inside of $\mathbb{C}^3$ intersecting at angle $\theta$. Na\"ively, one would say that the rotation matrix factorizes from the second determinant and one would end up with the same operator. However, in order to attach a string of letters to these operators, one has to cut the first determinant by deleting one factor of $\alpha X +\beta Y$, and then sew the result to the first letter of the word by matrix multiplication. In the planar limit, the corresponding boundary interactions for the spin chain will be
\begin{equation}
\begin{aligned}
  H_{\text{boundary}} = \lambda &\left(Q^{\alpha X +\beta Y}_0(1- P_{0,1}) Q^{\alpha X +\beta Y}_0 \right. \\ &\left. +\, Q^{\alpha e^{i\theta}X +\beta e^{-i\theta} Y}_{L+1}(1- P_{L,L+1})Q^{\alpha e^{i\theta}X +\beta e^{-i\theta} Y}_{L+1}\right).
\end{aligned}
\end{equation}
These interactions may be interpreted as turning the boundary sites into background $SO(6)$ magnetic fields. For clarity we may restrict to an $SU(3)$ holomorphic sector of the spin chain where $K_{i,i+1}=0$. In this sector the boundary fields polarize the first site of the spin chain in either the $Z$ or $\beta^* X -\alpha^* Y$ directions, and similarly for the last sites. Such boundary conditions are known to be integrable, and the corresponding solutions to the boundary Yang-Baxter equations are well studied \cite{Cao:2013cja}; for a textbook-level review see \cite{2015odba.book.....W}.

Geometrically, we expect that these interactions will implement mixed boundary conditions on the open strings localized near the intersection of the complex lines. The $Z$ direction of the geometry will correspond to the $U(1)$ fiber of the complex Hopf fibration of the five-sphere $S^5 \rightarrow \mathbb{CP}^2$. This is because the boundary conditions explicitly break the $SU(3)$ symmetry of the Hamiltonian down to $SU(2)\times U(1)$, where the $U(1)$ charge is associated to the number of $Z$'s in the spin chain. Since the projectors at the boundaries (the $Q$'s above) leave $Z$ unchanged, there is no way to change the number of $Z$'s without increasing the energy of the state. So we should expect that the size of the $U(1)$ fiber associated to $Z$ is fixed, and that the number of $Z$'s in the chain is the angular momentum along the fiber. In other words, we should expect to see that the geometry of the intersection of the holomorphic cycles inside $\mathbb{CP}^2$ is encoded in the spin chain.

\subsection{ABJM}
For the ABJM spin chain, the geometry of the intersection of the giant gravitons is more complicated since they are cycles inside of $\mathbb{CP}^3$ as opposed to an odd-dimensional sphere where one of the directions can be chosen along the circle fiber. For example, the operator
\begin{equation}
    \det \left(A_1 B_1 \right)
\end{equation}
describes a D4-brane wrapping two $\mathbb{CP}^2$'s of different orientations intersecting at a $\mathbb{CP}^1$ inside $\mathbb{CP}^3$. This can be understood in homogeneous coordinates, where we parametrize $\mathbb{C}^4$ in terms of a doublet of complex numbers $a_{1,2}$, $b_{1,2}$ and then projectivize. The corresponding locii are given by
\begin{equation} \label{brane1}
    \begin{aligned}
   a_1 b_1=0.
    \end{aligned}
\end{equation}
This equations defines a pair of $\mathbb{C}^3$'s intersecting at a single $\mathbb{C}^2$. We could also consider a different family of cycles with the same topology by replacing \eqref{brane1} with
\begin{equation}
    \mathcal{R}_A (a_1)     \mathcal{R}_B (b_1) =0,
\end{equation}
where $\mathcal{R}_{A,B}$ are distict $SU(2)$ matrices which rotate the $a$ and $b$ coordinates independently. After projectivization, this will define a different giant graviton---one which intersects the first one at angles given by the eigenvalues of $\mathcal{R}_{A,B}$. 

Using the same logic, one can consider a long operator $\mathcal{W}^I_J$ which attaches on one side to the determinant $ \det \left(A_1 B_1 \right)$ and to $ \det \left( \mathcal{R}_A (A_1)     \mathcal{R}_B (B_1)\right)$ on the other. Acting with the planar dilatation operator on such an operator would describe the Hamiltonian on an open spin chain with a particular set of reflecting boundary conditions. 

For a string that starts and ends on the same giant graviton, three distinct combinations of boundary conditions (and their orientation reversals) are allowed, depending on where the endpoints lie. For example, a string can start and end at the loci $a_1=0$. This would correspond to an operator of the following form:
\begin{equation}
\begin{aligned}
O_{\mathcal{W}}&= \epsilon_{i_1 \dots i_{N}  } \epsilon^{j_1 \dots j_N } (A_1 B_1)_{j_1}^{i_1} \dots(A_1 B_1)_{j_{N-1}}^{i_{N-1}}  \mathcal{W}_{j_N}^{i_N}, \\
\mathcal{W}&= A_2 B_{b_1} A_{a_2}\dots B_{b_L} A_{2}.
\end{aligned}
\end{equation}
One can also consider strings that start at $a_1=0$ and end at $b_1=0$; these are represented by operators where the first $A$ letter cannot be $A_1$ and the last $B$ letter cannot be $B_1$. 
A more interesting class of operators are those which correspond to strings which lie inside the $\mathbb{CP}^1$ intersections described by the equations $a_1=b_1=0$ and $  \mathcal{R}_A (a_1)=\mathcal{R}_B (b_1) =0$. These open strings have mixed boundary conditions, which make the boundary terms of the spin chain more complicated at both endpoints. Additionally, since the intersection of the two projective lines is non-trivial, the open spin chain will have non-trivial reflection matrices at the boundaries which encode information about this intersection. The corresponding boundary interactions are similar to the ones in the previous section:
\begin{equation}\label{boundary terms}
\begin{aligned}
H_{\text{boundary}}&= \frac{\lambda^2}{8 \pi^2} Q^{A_1}_1 Q^{B_1}_2\left( 1- P_{1,3}+ 1-P_{2,4}  \right)Q^{A_1}_1 Q^{B_1}_2 Q^{\mathcal{R}_A(A_1)}_{2L-1} Q^{\mathcal{R}_B(B_1)}_{2L}\\
 &\frac{\lambda^2}{8 \pi^2}  Q^{\mathcal{R}_A(A_1)}_{2L-1} Q^{\mathcal{R}_B(B_1)}_{2L}\left( 1- P_{2L-2,L}+ 1-P_{2L-3,2L-1}  \right)Q^{A_1}_1 Q^{B_1}_2 Q^{\mathcal{R}_A(A_1)}_{2L-1} Q^{\mathcal{R}_B(B_1)}_{2L}.
\end{aligned}
\end{equation}
Interestingly, after projecting out unphysical modes, (\ref{boundary terms}) can be thought of as the Hamiltonian for two Heisenberg chains of length $L-2$ coupled to boundary magnetic fields:
\begin{equation}
    Q^{\ket{\downarrow}} \left(1- P_{1,3}\right) Q^{\ket{\downarrow}}=\frac{1}{2}\left(1+ \sigma^z_1 \right) \sigma_2^z\sim \boldsymbol{h}\cdot \boldsymbol{\sigma}_2.
\end{equation}
This works because any state where the first spin is not aligned with the $z$-axis gets projected out by $Q^{\ket{\downarrow}}$, so that in practice we can treat the first site as fixed. This chain is known to be integrable for arbitrary choices of the boundary magnetic fields and the transfer matrix is explicitly known \cite{Cao:2013cja}.

\section{Open String Spectrum}

\subsection{$SU(3)$ sector of $\mathcal{N}=4$ SYM}

As we mentioned in the previous section, the boundary conditions leave boundary sites with $\ket{Z}$ states untouched on both ends. This means that the boundary conditions are triangular, in the sense that $\ket{Z}^{\otimes(L+1)}$ is the ground state of the system. This state also serves as a natural reference state for a coordinate Bethe ansatz. The main issue at hand is that there is no natural notion of defect due to the non-diagonal boundary conditions. That is to say that a plane wave of $X$ (or $Y$) defects might be absorbed and reflected as a plane wave of $Y$'s (or $X$'s, respectively) with some probability, so that the number of $X$'s or $Y$'s in the chain are not individually well-defined. This means that to find the excited states of the system, one has to consider possibly complicated combinations of $X$ and $Y$. 

\subsubsection{One-defect states}
Even though we know that the one-defect eigenstates are going to be linear combinations of plane waves with fixed momentum,\footnote{ This is because the system is integrable, so we should not expect that momentum is transferred at the boundaries.} it will soon become clear that the conditions for having an eigenvector of the Hamiltonian have a nice geometric interpretation. For example, consider the most general one-magnon state,
\begin{equation}
    \ket{\varphi}= \sum_{n=0}^{L+1}\left( x_n \ket{Z^n X Z^{L-n}}+  y_n \ket{Z^n Y Z^{L-n}}\right).
\end{equation}
The action of the bulk Hamiltonian (for sites between 1 and $L$) will act as a discrete second difference operator:
\begin{equation}
\label{recursion}
\begin{aligned}
    2x_{n}- x_{n+1}-x_{n-1}&= \varepsilon \,x_n\\
     2y_{n}- y_{n+1}-y_{n-1}&= \varepsilon\,  y_n.
\end{aligned}
\end{equation}
These conditions are essentially a discretized Schrodinger equation for a pair of wavefunctions $x_n, y_n$ with energy $\varepsilon$. One might think that this is simply a pair of decoupled second difference equations, but because the state at each site is normalized by $|x_n|^2+|y_n|^2=1$, the correct interpretation is that of a standing wave in $\mathbb{CP}^2$. The contributions to the eigenvalue equations from the boundary interactions are more complicated. For the zeroth site, these look like
\begin{equation}
\begin{aligned}
    \varepsilon\, x_0&= |\beta|^2 (x_0-x_1)- \alpha^* \beta (y_0-y_1),\\
     \varepsilon\, y_0&=-\beta^* \alpha (x_0-x_1)+ |\alpha|^2 (y_0-y_1),\\
      \varepsilon\, x_1&= 2x_1- x_2- |\beta|^2 x_0+ \alpha^*\beta y_0,\\
       \varepsilon\, y_1&= 2y_1-y_2 -|\alpha|^2 y_0 + \beta^* \alpha x_0.
\end{aligned}
\end{equation}
There are similar conditions that arise from the interactions at the last site, but in this form they are not very useful. Even though the conditions look complicated, they may be decoupled in a simple way. By taking linear combinations of the first two equations, we obtain the consistency condition
\begin{equation}
    \alpha x_0+ \beta y_0=0, \qquad \varepsilon \neq 0.
\end{equation}
The same can be done for the other end of the spin chain; the corresponding constraint being
\begin{equation}
    \alpha e^{i\theta} x_{L+1}+\beta e^{-i\theta} y_{L+1}=0, \qquad \varepsilon \neq 0.
\end{equation}
In other words, the one defect wavefunctions satisfy two different Dirichlet boundary conditions at each endpoint. Geometrically, this means that the spin chain state ``ends'' on a pair of holomorphic cycles sitting inside the five-sphere $|x|^2+|y|^2+|z|^2=1$ given by $\alpha x_0+ \beta y_0=0$ and $ \alpha e^{i\theta} x_{L+1}+\beta e^{-i\theta} y_{L+1}=0$. So indeed, the operators $\mathcal{O}_{0, L+1}$ describe D-branes wrapping those cycles, and the spin chain state describes an open string stretching between them. Once these conditions are realized, it is not hard to see that the second pair of equations can be turned into the second difference equations
\begin{equation}
 \begin{aligned}
      \varepsilon\, x_1&= 2x_1- x_2-  x_0+ \alpha^*(\alpha x_0+\beta y_0),\\
       \varepsilon\, y_1&= 2y_1-y_2 -y_0 + \beta^* (\alpha x_0+\beta y_0).\\
\end{aligned}   
\end{equation}
After some more algebraic manipulations, the remaining Neumann boundary conditions are 
\begin{equation}
\begin{aligned}
      \beta^*(x_1-x_0)- \alpha^*(y_1-y_0)&= -\frac{\varepsilon \,x_0}{\beta}\\
      e^{i\theta} \beta^*(x_{L+1}-x_{L})-  e^{-i\theta}\alpha^*(y_{L+1}-y_{L})&= \frac{e^{i\theta}\varepsilon \,x_{L+1}}{\beta}.
\end{aligned}
\end{equation}
To interpret these boundary conditions, we should remember that the probability current associated to a wavefunction is given by
\begin{equation}
    \boldsymbol{j}= \Im\Psi^* \boldsymbol{\partial} \Psi.
\end{equation}
For simple Neumann boundary conditions on the wavefunction $\frac{\partial \Psi}{\partial n}\big|_{B}=0$, this says that the flux of $\boldsymbol{j}$ will vanish at the the Dirchlet boundary, i.e, $(\boldsymbol{n}\cdot \boldsymbol{j})\big|_{B}=0$. For more general mixed boundary conditions, this need not be the case: as long as the net flux is conserved, the Dirichlet boundaries may exchange charge. Since the right-hand side of both Neumann boundary conditions is a constant, as long as their sum vanishes the corresponding wavefunction is self-consistent:
\begin{equation}
\begin{aligned}
         x_0&=e^{i \theta} x_{L+1},\\
         y_0&= e^{-i \theta} y_{L+1}.
\end{aligned}
\end{equation}
Since the solution to the recursion relations (\ref{recursion}) is $x_n= A_X e^{i\varphi n}+ B_Xe^{-i\varphi n}$, the Neumann boundary condition imposes a quantization condition on the quasi-momentum $\varphi$:
\begin{equation}
    \varphi \pm \theta = \frac{2\pi k}{L+1}=p.
\end{equation}
The dispersion relation $\epsilon(p)$ is similar to the case with periodic boundary conditions:
\begin{equation}
    \varepsilon(p)_{\pm}= \frac{ \lambda}{8 \pi^2}\left(2- 2\cos\left(p \pm \frac{\theta}{L+1} \right)\right)= \frac{ \lambda}{2 \pi^2 } \sin^2 \frac{1}{2}\left(p \pm \frac{\theta}{L+1} \right)
\end{equation}
Since the phase shift is different for $x_n$ and $y_n$, the eigenstates have to be complicated combinations of left and right moving waves of $X$ and $Y$ with fixed momentum:
\begin{equation} \label{wavefnc}
\begin{aligned}
 \ket{k}&=\sum_{n=0}^{L+1}\,\left( A_- e^{i\left(\frac{2\pi k}{L+1}- \frac{\theta}{L+1} \right)n} +A_+ e^{i\left(\frac{2\pi k}{L+1}+ \frac{\theta}{L+1} \right)n} \right)\ket{Z^n X Z^{L-n}} \\
 &+\left( B_- e^{i\left(\frac{2\pi k}{L+1}- \frac{\theta}{L+1} \right)n} +B_+ e^{i\left(\frac{2\pi k}{L+1}+ \frac{\theta}{L+1} \right)n} \right) \ket{Z^n Y Z^{L-n}}.
\end{aligned}
\end{equation}
So the effect on one-defect states of including this class of non-diagonal boundary conditions is to shift the quantization of the momentum by a factor of $\frac{\theta}{L+1}$. In practice, solving for the coefficients $A_{X,Y}, B_{X,Y}$ for generic $\theta$ involves solving very complicated algebraic equations.

It should be noted that this is precisely the mode expansion of the worldsheet fields for an open string near the intersection of two D-branes which intersect at an angle $\theta$ \cite{Berkooz:1996km}:
\begin{equation}\label{berkooz}
     X^I(w, \bar{w}) =\sum_{k\in \mathbf{Z}}\left( x_{k- \frac{\theta}{L+1}}^I e^{i\left(\frac{2\pi k}{L+1}- \frac{\theta}{L+1} \right)w}+\tilde{x}_{k+\frac{\theta}{L+1}}^Ie^{-i\left(\frac{2\pi k}{L+1}+ \frac{\theta}{L+1} \right)\bar{w}}\right).
\end{equation}

To see that this is the correct matching, we can first consider the plane-wave limit $L\sim \sqrt{N}$, where the dispersion relation in units of the light-cone momentum $p^+\sim L\sim J$ is given by
\begin{equation}
   \left( \Delta - J\right)_k= 1+ \frac{g^2 N (k\pm \frac{\theta}{2\pi})^2}{2 J^2}+ \dots,
\end{equation}
which agrees precisely the BMN scaling for the energy on the plane-wave with a momentum shift of $\frac{\theta}{2\pi}$ :
\begin{equation}
    H_{lc}= \sum_{n=-\infty}^\infty N_n \sqrt{\mu^2+ \frac{ g^2 N(n\pm \frac{\theta}{2\pi})^2}{ J^2}},
\end{equation}
where $\mu$ can be re-scaled to one. In this limit, the dispersion relation \eqref{dispersion} reduces to a relativistic dispersion relation of a massive particle with unit mass in $AdS$ units. Since the spin chain interactions are suppressed, the system reduces to a tower of relativistic massive bosons as one would expect from the string spectrum.  One can then recover the flat-space spectrum by taking the massless limit $\mu \rightarrow 0$ reproducing \eqref{berkooz}.

For states with multiple defects at finite volume, one would expect that the boundary conditions modify the Bethe equations for the momenta in an appropriate way. Since the Hamiltonian in the bulk of the spin chain is the same for periodic and open boundary conditions, the two-magnon S-matrix is the same except that the momentum should be shifted for open boundaries. The main complication is that the Bethe equations must be modified at the boundaries.  In principle, this modification can be very complicated \cite{Zhang:2015fea}: it would encode the first curvature corrections to the spectrum of open strings between the giant gravitons at angles. One should also be able take into account the effect of wrapping corrections \cite{Bajnok:2015kfz} on the wavefunction \eqref{wavefnc}. The study of such states is beyond the scope of this work. 

\subsection{$SU(2)\times SU(2)$ sector of ABJM}

Since the two $SU(2)$ factors of this sector do not interact with each other, we can consider a single $XXX$ chain with non-diagonal boundary conditions. This spin chain serves as a good example of a more complicated boundary condition because, unlike in the previous section, now there is no obvious reference state. This is due to boundary conditions which do not allow for a ground state with a $U(1)$ symmetry. A general choice of determinant operators will give a pair of boundary reflection matrices that cannot be simultaneously diagonalized. In some cases, the boundary conditions may be cast into a form where one of the boundary magnetic fields is fixed in, say, the $z$ direction, while the other boundary interaction is some upper-triangular matrix. We will not restrict ourselves to this case, as it is identical to the analysis of the previous section and there is a clear way of implementing a coordinate Bethe ansatz. The reference state in those cases will correspond to a rotating point-like string with light-cone momentum $Q=L+1$. Generically, we will have a situation where only one of the boundary interactions can be put in diagonal form, while the other is not upper-triangular. Here there is no clear ground state, which means that we should expect to see the open string being dragged in different directions by the giant gravitons at the endpoints. This also means that the would-be $U(1)$ symmetry is gauged along the string; the ground state must carry a non-trivial central charge from one endpoint to another. 

\subsection{Approximating the ground state}

Finding the exact ground state for the spin chain with non-diagonal boundaries is difficult for finite-size spin chains. One way to approximate this state is by using a variational wavefunction made out of spin-coherent states \cite{Kruczenski:2003gt}.  One reason to expect that this is a good starting point is that one expects that the ground state has a non-trivial magnetic field gradient. For completeness, the variational ansatz in question is given by \cite{Stefanski:2004cw}:
\begin{equation}
\begin{aligned}
   \ket{\Psi}&= \bigotimes_{i=0}^{L+1} \ket{\Vec{\alpha}_i}, \qquad
     \ket{\Vec{\alpha}}= \exp{\frac{i}{2}\left(n_x \sigma^x- n_y \sigma^y \right)} \ket{\uparrow}, \qquad |\Vec{\alpha}_i|^2=1.
\end{aligned}
\end{equation}
The vector $\vec{\alpha}$ gives coordinates for the coset $SU(2)/U(1)$, and is related to $\Vec{n}$ as follows:
\begin{equation}
\begin{aligned}
        \alpha_x &= n_x \frac{\sin \Delta}{\Delta}, \qquad
     \alpha_y = n_y \frac{\sin \Delta}{\Delta}, \qquad
     \alpha_z= \cos \Delta, \qquad
     \Delta= \sqrt{n_x^2 +n_y^2}.
\end{aligned}
\end{equation}
These auxiliary coordinates are useful because the energy evaluates to a sum of squares:
\begin{equation}
\label{hamiltonian-exp}
    \langle H \rangle= \frac{\lambda^2}{8 \pi^2}\sum_{i=0}^L |\Vec{\alpha}_{i+1}- \Vec{\alpha}_i|^2.
\end{equation}
To minimize the energy, one has to minimize (\ref{hamiltonian-exp}) subject to the constraint that the $\Vec{\alpha}$ are normalized, which can be implemented by introducing a Lagrange multiplier $\Lambda$:
\begin{equation} \label{coherent ham}
    E= \frac{\lambda^2}{8 \pi^2}\sum_{i=0}^L |\Vec{\alpha}_{i+1}- \Vec{\alpha}_i|^2 -\frac{\lambda^2}{8 \pi^2} \sum_{i=0}^{L+1}\Lambda(|\Vec{\alpha}_i|^2-1).
\end{equation}
This function is extremized when the bulk parameters $\Vec{\alpha}_i$ satisfy the following second difference eigenvalue equation:
\begin{equation}
    2 \Vec{\alpha}_i-  \Vec{\alpha}_{i-1}- \Vec{\alpha}_{i+1}= \Lambda  \Vec{\alpha}_i.
\end{equation}
The Lagrange multiplier can be solved for after normalizing the $\Vec{\alpha}$; the result is
\begin{equation}
    \Lambda= 2- \Vec{\alpha}_{i-1}\cdot \Vec{\alpha}_{i}- \Vec{\alpha}_{i+1}\cdot \Vec{\alpha}_{i}.
\end{equation}
Notice that these equations are the same as the Bethe ansatz for a one-magnon state, except that the wavefunction lives on the sphere $SU(2)/U(1)$. In the continuum limit, these equations are exactly the equations of motion for a stationary open string moving inside a two-sphere. Because $\Lambda$ satisfies the same equation for every $i$, the angle between any two neighboring spins is the same:
\begin{equation}
   \Vec{\alpha}_{i+1}\cdot \Vec{\alpha}_{i} \equiv \cos \left(\frac{\Phi}{L+1}\right).
\end{equation}
This immediately gives us the form of the energy of this variational state:
\begin{equation}
    \langle H\rangle =2(L+1)\frac{\lambda^2}{8 \pi^2}\left[1-\cos \left(\frac{\Phi}{L+1}\right)  \right].
\end{equation}
This suggests that to leading order, the true ground state can be viewed as a superposition of $L+1$ magnons with momenta $p= \frac{\Phi}{L+1}$. The angle $\Phi$ is simply the angle between the boundary parameters $\Vec{\alpha}_0\cdot \Vec{\alpha}_{L+1}= \cos \Phi$ in the coset coordinates. This angle is not the same as the geometric angle between the background magnetic fields (i.e. the angle at which the D-branes intersect). To find this latter angle, we can align $\Vec{n}_0$ with the $z$-axis, and set $\Vec{n}_{L+1}$ along the $x$-$y$ plane at an angle $\theta$ to the $x$-axis.  In terms of the geometric angle $\theta$ between $\Vec{n}_0$ and $\Vec{n}_{L+1}$, the energy has a more complicated form:
\begin{equation}
    \langle H\rangle = 2(L+1)\frac{\lambda^2}{8 \pi^2}\left[1-\cos \left(\frac{\sin\theta}{L+1}\right) \right]\approx \frac{\lambda^2}{2 \pi^2(L+1)} \sin^2 \theta+ \mathcal{O}(1/L^2).
\end{equation}
In the thermodynamic limit, this agrees with the expected form of the energy coming from the dispersion relation of a centrally-extended BPS state:
\begin{equation}
    \sqrt{(L+1)^2+ \frac{\lambda^2}{\pi^2}\sin^2 \theta}\approx (L+1)+ \frac{\lambda^2}{2\pi^2(L+1) } \sin^2 \theta+ \dots.
\end{equation}
This means that the while the semiclassical coherent state ansatz captures the correct physics for large chains $L\gg1$, and it must receive quantum corrections for small chains.

In the limit of large angular momentum $J\rightarrow \infty$, the classical sigma model in the $SU(2)\times SU(2)$ sector is equivalent to a pair of Landau-Lifshitz models with a level matching constraint \cite{Grignani:2008is}:
\begin{equation}
\begin{aligned}
     S&= \sum_{i=1}^2\int d\tau \int d\sigma \left[y_i \dot{x}_i -\frac{ \lambda^2}{4 \pi^2 J^2}\left[\left( x_i'^2+y_i'^2+z_i'^2\right)- \Lambda\left( x_i^2+y_i^2+z_i^2-1\right)\right]\right]
\end{aligned}
\end{equation}
This is simply a sigma model whose target space is a pair of spheres. As before, we can concentrate on one of the spheres, since the other contribution to the action can be seen as an additional open string whose end points lie on the other sphere. Since the Lagrangian is already of the form $L=q\dot{p}- H$, it is clear that the classical Hamiltonian for time independent solutions the action reduces to:
\begin{equation}
     H= \frac{ \lambda^2}{4 \pi^2 J^2} \sum_{i=1}^2 \int_o^J d\sigma \left[\left( x_i'^2+y_i'^2+z_i'^2\right)- \Lambda\left( x_i^2+y_i^2+z_i^2-1\right)\right],
\end{equation}
which is simply the continuum limit of the coherent state energy functional \eqref{coherent ham}. Solving the equations of motion and imposing the boundary conditions as before gives the following expression for the action:
\begin{equation}
   \frac{ \lambda^2}{4 \pi^2 J^2} \int_0^J d\sigma \Lambda= \frac{ \lambda^2}{4 \pi^2 J^2}\int_0^J d\sigma (\partial x_i)^2= \frac{\lambda^2 \theta^2}{ 2 \pi^2 J}\sim \frac{ \lambda^2}{4 \pi^2 J^2}\left[2-  2\cos \left( \theta/J\right)\right],
\end{equation}
which is the long wavelength behavior of the spin chain. This makes sense, since in the limit of large angular momentum the geodesic motion of the string is pushed towards the intersection loci of the D-branes. Note that in this case the resulting limit is different from  the plane-wave limit, since the target space is $AdS_4\times \mathbb{CP}^3$, but the high angular velocity of the string forces the worldsheet towards the intersection locus of the D4 giant gravitons. These are the rotating string solutions studied in \cite{Kruczenski:2003gt} and the corresponding infrared dynamics of the corresponding spin chain.
\subsubsection{Approximate excited states}
In the previous section we saw that the ground state for the $XXX$ chain with non-diagonal boundaries carries a non-trivial central charge which encodes the angle of intersection of a pair of giant gravitons associated to each boundary condition. At large $L$, the coherent state is an approximate eigenvector of the Hamiltonian:
\begin{equation}
    H  \ket{\Psi(\theta)}\approx E \ket{\Psi(\theta)}+ \mathcal{O}(1/L).
\end{equation}
As such, we may treat it as an approximate reference state and attempt a coordinate Bethe ansatz. For example, a one-defect approximate state is of the form
\begin{equation}
    \ket{p}\approx \sum_{l=1}^{L} F_l e^{i p l}\ket{\Psi(\theta)},
\end{equation}
where $F_l$ is an operator that exchanges $\Vec{\alpha_l}$ with $-\Vec{\alpha_l}$. When one acts on this state with the Hamiltonian, the leading-order energy will roughly be of the form
\begin{equation}
    H\ket{p}\approx \lambda^2\left[2-2\cos\left(p+\frac{\Phi}{L+1} \right)\right]\ket{p} + \mathcal{O}(1/L).
\end{equation}
In this dilute gas approximation, the momentum $p$ is also quantized since the wavefunction will resemble a standing wave of wavelength $\frac{2 \pi k}{L+1}$.

\section{Discussion}

We have considered a general class of boundary conditions for integrable spin chains and showed that one is able to reproduce various aspects of open strings stretching between intersecting $D$-branes. Even though we concentrated on simple examples, the lessons here should apply generally for all maximal giant gravitons. One should expect that a general rule for producing boundary conditions is to introduce projector operators on the boundary sites, or equivalently to turn the boundary sites into fixed sources. We saw that for $SU(k)$ spin chains with $k>2$, one can usually fix the ground state by choosing a reference state that is left fixed by the boundary interactions. This condition should correspond to a statement about the supersymmetry being preserved by each of the boundaries of the string. In other words, even though both giant gravitons preserve one-quarter of the supersymmetry algebra, depending on the nature of the intersection, this can be enhanced to one-half of the supersymmetries. After fixing the reference state, the one-defect states are a particular combination of plane-waves whose momentum is shifted by an amount that encodes the angle of the intersection of the giant gravitons, as one would expect from the local analysis near the intersection where the geometry looks flat. In principle, one should be able to take superpositions of such states satisfying appropriate boundary conditions to find a set of modified Bethe equations that fix the quantization of the momenta at each level. Since much of the analysis is parallel to the case of periodic boundary conditions, one should expect that the full open string spectrum between maximal giants that preserve mutual supersymmetries can be obtained by solving the appropriately modified non-perturbative Bethe equations along the lines of \cite{Zhang:2015fea} and \cite{Bajnok:2013wsa}. In particular, it would be interesting to obtain explicit expressions for the magnon momenta as a function of the angle $\theta$ between the branes. 

Another possibility is that the boundary interactions do not share a mutual eigenvector; in this case one expects that there is no supersymmetry enhancement at the intersection, and the ground state is allowed to carry a non-trivial central charge. This is different from the analysis in flat space, where intersecting D-branes always have enhanced supersymmetry at their intersections. This effect is due to the fact that giant gravitons are stabilized by angular momentum or flux, and one can have situations where the giants do not rotate along a mutual circle, or where the fluxes stabilizing them are not parallel or perpendicular.  We studied one instance of this possibility in the ABJM spin chain since the boundary interactions of that system are more restricted. In this case the ground state has a non-trivial zero-point energy associated to the central charge being carried along the string. Spin chains whose ground states carry a non-trivial central charge have been studied before, but only in the case where the central charge arises from having non-maximal giants which are parallel and non-intersecting. Although we were unable to find the explicit ground state of the system, we used a variational ansatz to obtain an upper bound on the ground state energy which happened to follow the correct behavior expected for a rotating string. It would be interesting to see whether the ground state itself can be found explicitly. One way of approaching this would be to try to introduce a gauge field $a_i$ along the chain which mediates the spontaneously broken $U(1)$ charge. The ground state should correspond to a state with all spins aligned in one direction dressed with a Wilson line for $a_i$. It would be interesting to see if there is an appropriate notion  of a \textit{gauged} coordinate Bethe ansatz, where the coordinate Bethe ansatz is dressed appropriately to take into account the flow of central charge. One should be also be able to capture properties of this ground state and low-lying states by finding the corresponding sigma model classical solutions.

The cases with non-maximal giants are complicated by the fact that the spin chain becomes dynamical, meaning that the number of sites is no longer a good quantum number for the system. One way of dealing with this issue is to fix the number of times a particular field appears in a gauge invariant word, and treat all other fields as defects. This would correspond to a particular Neumann boundary condition along the string, where the size of the circle associated to the angular momentum $L+1$ is allowed to vary. These kinds of spin chains have excitations with Boltzmann statistics, since the order in which the matrices are multiplied in the gauge invariant operator matters. Once one has a description of these types of system in terms of these ``free'' bosons, one may study generic boundary conditions for such a chain in which more complicated combinations of letters can hop on and off the spin chain. We expect that those boundary conditions describe more complicated configurations of giants wrapping holomorphic cycles \cite{Mikhailov:2000ya}.It would also be interesting to study the corresponding open string solitons with those open boundary conditions. One should be able to reproduce the quantum numbers of the spin chain in the limits studied in \cite{Hofman:2006xt, Berenstein:2020jen}. These are different than the plane-wave and Landau-Lifshitz limits in that they capture some finite size effects in the dispersion relation. 

Another important class of boundary conditions that we have not considered in this work, occur when the boundary interactions are not mutually holomorphic (i.e. where states with $X$ on the first site  and $\bar{X}$ are projected out). These correspond to non-supersymmetric brane-antibrane configurations. The bosonic spectrum with such boundary interaction is expected to contain a tachyon corresponding to the a decay channel for the branes \cite{Bajnok:2013wsa}. Since we can begin with intersecting brane systems where the angle between the brane and antibrane is small, one can consider double scaling limits where the t'Hooft coupling is taken to be large, but the effective expansion parameter is taken to be small ($\frac{\lambda \sin^2 \theta}{L+1}\lll 1$) so that perturbation theory is applicable on both sides of the duality. This provides another avenue for studying unstable brane configurations in curved spaces. These are issues under current investigation.

Our analysis should apply more generally and it would be interesting to understand which open spin chains also admit a geometric description in terms of rotating open strings stretching between D-branes in curved spaces. If the so-called ``string hypothesis'' is true in some form, then one can expect that all integrable open chains have such a description. Cases of particular interest also include the non-integrable spin chains arising from more realistic gauge theories such as large $N$ QCD \cite{Beisert:2004fv}, Lagrangian superconformal theories with less supersymmetry \cite{Pomoni:2021pbj}, and fishnet theories \cite{Gurdogan:2015csr, Ahn:2020zly}. Our methods should provide a starting point to probe the geometry of the bulk string dual of other large $N$ gauge theories. 


\acknowledgments
We thank David Berenstein for the many fruitful discussions and suggestions regarding this project. We especially thank Alex Meiburg for the many insightful discussions we had during the preparation of this work.  We also thank David Grabovsky and Wayne Weng for comments on this manuscript.

\end{document}